# The Physical Foundation of the Reconnection Electric Field

M. Hesse[1,2], Y.-H. Liu[3], L.-J. Chen[4], N. Bessho[4], S. Wang[4], J. Burch[2], T. Moretto[1], C. Norgren[1], K. J. Genestreti[5], T. D. Phan[6], P. Tenfjord[1]

[1]Birkeland Centre for Space Science, Department of Physics and Technology, University of Bergen, Norway. [2]Southwest Research Institute, San Antonio, TX, USA. [3]Dartmouth College, Hanover, NH, USA. [4]NASA, Goddard Space Flight Center, Greenbelt, MD, USA. [5]Space Research Institute, Austrian Academy of Sciences, Graz, Austria. [6]University of California, Berkeley, CA, USA.

**ABSTRACT**

Magnetic reconnection is a key charged particle transport and energy conversion process in environments ranging from astrophysical systems to laboratory plasmas[1]. Magnetic reconnection facilitates plasma transport by establishing new connections of magnetic flux tubes, and it converts, often explosively, energy stored in the magnetic field to kinetic energy of charged particles[2]. The intensity of the magnetic reconnection process is measured by the reconnection electric field, which regulates the rate of flux tube connectivity changes. The change of magnetic connectivity occurs in the current layer of the diffusion zone, where the plasma transport is decoupled from the transport of magnetic flux. Here we report on computer simulations and analytic theory to provide a self-consistent understanding of the role of the reconnection electric field, which extends substantially beyond the simple change of magnetic connections. Rather, we find that the reconnection electric field is essential to maintaining the current density in the diffusion region, which would otherwise be dissipated by a set of processes. Natural candidates for current dissipation are the average convection of current carriers away from the reconnection region by the outflow of accelerated particles, or the average rotation of the current density by the magnetic field reversal in the vicinity. Instead, we show here that the current dissipation is the result of thermal effects, underlying the statistical interaction of current-carrying particles with the adjacent magnetic field. We find that this interaction serves to redirect the directed acceleration of the reconnection electric field to thermal motion. This thermalization manifests itself in form of quasi-viscous terms in the thermal energy balance of the current layer. These quasi-viscous terms act to increase the average thermal energy. Our predictions regarding current and thermal energy balance are readily amenable to exploration in the laboratory or by satellite missions, in particular, by NASA's Magnetospheric Multiscale mission.

# I. INTRODUCTION

Magnetic reconnection involves an electron-scale current layer in the inner reconnection diffusion region[3]. This current layer separates the reconnection inflow region, where stored magnetic energy is available for conversion into other forms of energy. The current layer is bordered on the sides by the outflow region, with weaker magnetic fields created by the reconnection process, and where bulk flows transport plasma away from the region. This dynamic is the result of a complex balance. In fact, we show that processes conspire to reduce both current density and pressure, which would lead to the collapse of the layer if not for the energization by the local electric field.

The electric field in this region, referred to as the reconnection electric field, determines the creation of magnetic flux in the outflow region from magnetic flux in the inflow region by virtue of Faraday's law[4]. It thereby determines the creation of new magnetic connections, and, indirectly, regulates the energy conversion from inflow to outflow. This view, however, does not capture the complexity of the physical reason for the existence of the reconnection electric field. Using high-resolution kinetic simulations, we demonstrate that this electric field in fact plays a key role in sustaining the current layer in the inner diffusion region. Furthermore we find that its very existence results from the physical requirements to provide pressure- and, more directly, current balance, in this layer.

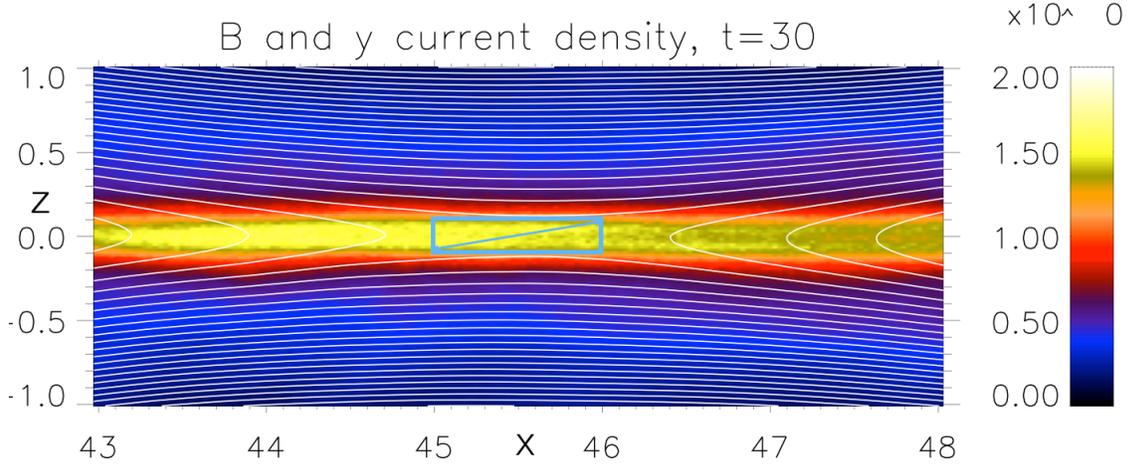

Figure 1. The integration region centered on the X point. Both current and internal energy conservation equations are integrated over a rectangular region of the form shown here. The rectangle shown here is of the largest size, and all rectangles are centered on the X point.

**II. MODEL**

Particle-in-cell simulations were performed in a 2.5 dimensional geometry similar to prior studies[5], using our well-proven simulation code. All quantities are presented in normalized form. The magnetic field is normalized to a typical, asymptotic, value $B_0$ of the reconnecting magnetic field, and the time unit is the inverse of the ion cyclotron frequency $\Omega_i = eB_0/m_i$. Velocities are measured in terms of the Alfven speed $v_A = B_0/\sqrt{\mu_0 m_i n_0}$, where $n_0$ is a typical value of the initial current layer density. The length scale is the ion inertial length $d_i = c/\omega_{pi}$ based on $n_0$. The simulation employs a ratio of electron plasma to electron cyclotron frequency of $\omega_{pe}/\Omega_e = 2$.

The initial configuration consists of a Harris sheet, $B_x = \tanh(z/l)$, where the current layer width is $l = 0.5 d_i$, with a small X-type, magnetic perturbation. Such, symmetric and antiparallel, configurations are typical for reconnection in the Earth's magnetotail. The ion/electron mass ratio is 100, and the dimensions of the simulated system are $L_x \times L_z = 102.4 d_i \times 51.2 d_i$. The simulation is performed on a 3200x3200 grid. A total of $7 \times 10^{10}$ particles are used. Shown in Figure 1 are magnetic field lines (white), and out-of-plane current density (y direction) in a segment of the larger simulation region. The blue rectangle denotes the largest region, over which we integrate current and energy equations (see below). The box size is varied from across a range of size, while keeping the aspect ratio of five fixed.

### III. CURRENT BALANCE

We use the high-resolution particle-in-cell simulation model described above to investigate both current and thermal energy balance. We choose to investigate a time during which reconnection is steady, and the reconnection rate $E_r = 0.17$ is in the typical range. The physics of the inner diffusion region is determined by the electrons[5]. We therefore investigate the electron dynamics. Assuming weak gradients in the current (y) direction, the balance equation for the out-of-plane electron current is:

$$-\frac{\partial}{\partial t} e n_e v_{ey} = \frac{e^2 n_e}{m_e} E_y + \frac{e^2 n_e}{m_e} v_{ez} B_x - \frac{e^2 n_e}{m_e} v_{ex} B_z + \frac{e}{m_e}\left(\frac{\partial P_{yze}}{\partial z} + \frac{\partial P_{xye}}{\partial x}\right) + e \nabla \cdot n_e \vec{v}_e v_{ey} \quad (1)$$

The terms on the right-hand-side describe, in order: acceleration by the electric field (first term), the conversion of in-plane current into out-of-plane current by the magnetic field (second and third terms, "Lorentz force"), the nonisotropic pressure force (fourth term in parentheses), and current changes by convection and compression (fifth term). Since (1) is the y-component of the momentum equation, the y-component of the electron velocity does not appear on the right-hand-side. This means that the deviation of the electron current by the magnetic field adjacent to the X point is a second order effect. In first order, only the interaction of the thermal motion of particles with the magnetic field around the X point can be important (see below).

We study this balance by integrating the terms of (1) over a rectangle, centered on the reconnection X point. The aspect ratio of the rectangle is five. In order to investigate the dependence of the results on the size of the integration region, we vary the half-thickness d of the rectangle shown in Fig. 1 across a broad range. Figure 2 shows the result. The nonvanishing integral that is seen for the time derivative of the current density is indicative of the fact that the system is not quite time-stationary.

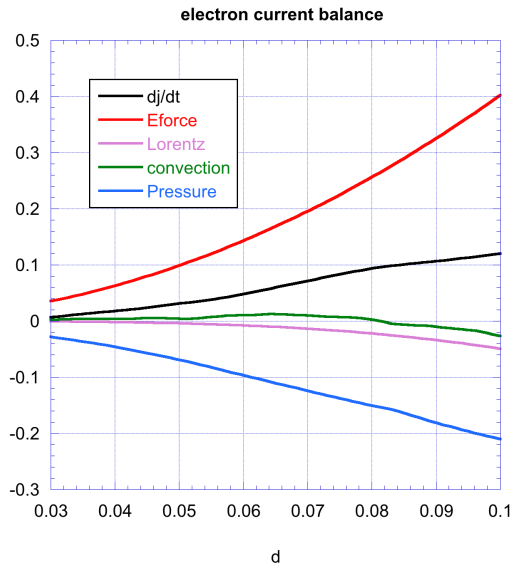

Figure 2. Integration of the electron current balance equation. The integrated time derivative of the out-of-place current density is indicated by the black graph, which is also equal to the sum of the four other graphs in agreement with eqn. (1). It shows a positive value for the current density derivative, indicative of a slow time dependence of the system. The convection term (green) is obtained by calculating the momentum flux

across all four boundaries of the integration box. Like the Lorentz force term (purple), it is small. The major contributors are the current increase by the reconnection electric field (red), and the current reduction by nongyrotropic (see below) pressure effects (blue). The dominance of these two terms holds over the entire range rather than at the X point alone.

The balance between electric field acceleration and pressure term is not unexpected at the X point proper[6]. However, we see here that its dominance extends over the entire range investigated here. We therefore find that thermal effects, associated with complex pressure and the interaction with the adjacent magnetic field gradients, are the main current reduction mechanism. This reduction is compensated for to match the time evolution of the current density, by the acceleration provided by the electric field. The electric field therefore plays the essential role of maintaining a current density, which would otherwise get reduced by effects primarily associated with thermal particle motion.

This pressure force is essentially exclusively due to nongyrotropic effects (Fig. 3), implying that particle distributions lack symmetry about the magnetic field. Nongyrotropic distributions are generated by the average, "thermal," interaction of current sheet particles with the adjacent magnetic field, one manifestation of which is the formation of crescent distributions[8-14,18]. Here we see that this type of interaction acts as the main current dissipation mechanism.

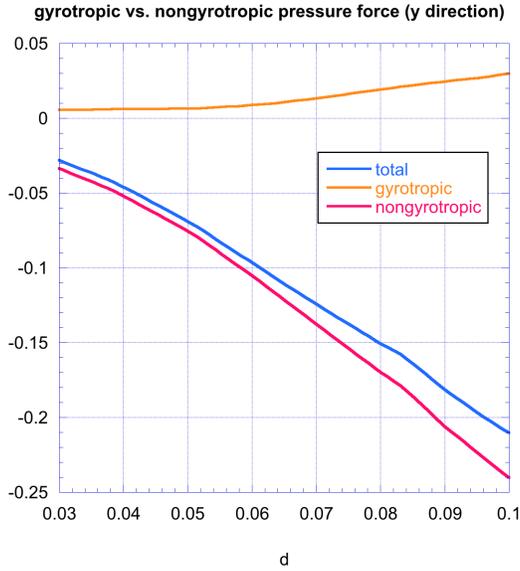

Figure 3. Decomposition of the pressure force into gyrotropic and nongyrotropic contributions. Nongyrotropic pressure is an indication of the lack of or only partial particle magnetization by the magnetic field. In a symmetric reconnection geometry, it can provide the main contribution to the reconnection electric field at the X point[5,7,15-17]. We here see that this is generically so, even in an extended region around the X point. The effects of the gyrotropic part of the pressure (orange) are negligible compared to those of the nongyrotropic part. Nongyrotropic pressures are created by the statistical interaction of current sheet particles with the ambient magnetic field, generating a current reduction mechanism similar to viscosity in a collisional plasma.

## III. ENERGY BALANCE

Intuitively, we would expect that the reduction of current flow by a process associated with thermal motion, should involve thermalization of the kinetic energy of the bulk flow supporting the current density. This thermalization should lead to an increase of the average, thermal, kinetic energy, or, equivalently, the scalar electron pressure. We will investigate this expectation now. The evolution of the pressure, defined by $p = Tr(\vec{P})/3$ derived from the trace of the tensor equation[19]. It is given by:

$$\frac{\partial p}{\partial t} = -\nabla \cdot (\vec{v}p) - \frac{2}{3}\sum_{l} P_{ll}\frac{\partial}{\partial x_l}v_l - \frac{1}{3}\sum_{l,i} \frac{\partial}{\partial x_i}Q_{lii} - \frac{2}{3}\sum_{\substack{l,i \\ l \neq i}} P_{li}\frac{\partial}{\partial x_i}v_l \qquad (2)$$

The first two terms on the right-hand-side of (2) represent pressure changes due to compression, expansion, and convection accounting for anisotropies in the plasma. The third term, given by the trace of the heatflux 3-tensor, constitutes a correction to the first two terms (see the appendix) for certain asymmetries in the distribution functions. As the highest order moment, the heat flux 3-tensor is most susceptible to statistical variations caused by limited particle numbers. Because it emphasizes particles of higher kinetic energy, it also requires a particularly accurate integration of particle orbits to properly account for the interaction of these particles with magnetic field variations. We took three measures in order to obtain a reasonably accurate representation of heat flux contributions. First, our simulation typically involves about 10000 particles per cell in the reconnecting layer. This huge number is necessary to increase the separation of numerical noise from physical signal so that higher order moments can be calculated without excessive averaging. Second, starting at $\omega_{pe}t=5979$, we reduced the time step, which heretofore had been $\omega_{pe}dt=0.25$, to $\omega_{pe}dt=0.01$, and ran the code until the end at $\omega_{pe}t=6000$, corresponding to $\Omega_i t=30$. This measure increases the critical accuracy of the electron orbit integration in the model. Outputs were produced for every full plasma period. Third, the heat flux was then averaged over outputs produced at full plasma period intervals between $\omega_{pe}t=5980$ and $\omega_{pe}t=5990$. For lower order moments, averaging over outputs produced at full plasma period intervals between $\omega_{pe}t=5983$ and $\omega_{pe}t=5987$ was sufficient.

Finally, the last term in (2) involves cross-derivatives of the flow velocity, which resemble viscous contributions in a collisional system, and off-diagonal pressure terms of the kind involved in current dissipation. We refer to this term as "quasi-viscous" and, based on the above discussion, expect it to lead to a pressure increase.

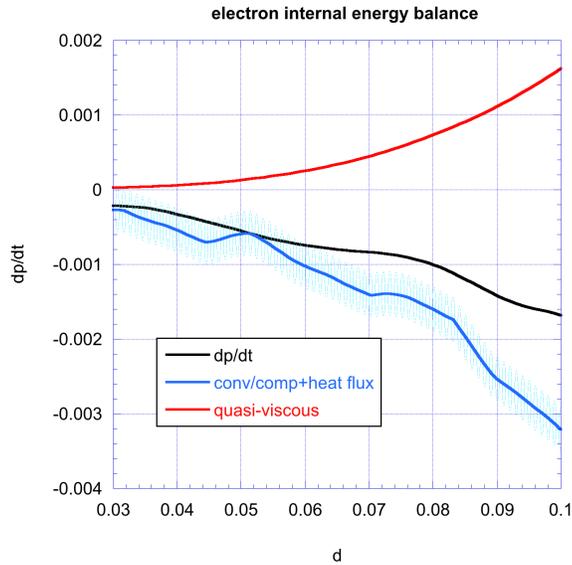

Figure 4. Electron internal energy balance. Shown are the spatially integrated time derivative of the pressure (black), the quasi-viscous heating term (red), and the combination of convection, compression, and heat flux (blue). The latter two graphs add up to the time derivative of the pressure within the error indicated by the shading of the blue graph. This error is attributable to the highest order moment, the heat flux. We see here that there is a slow pressure reduction due to the overall dynamics. In the absence of heating, this reduction would be considerably stronger. However, there is a significant

positive contribution due to quasi-viscous heating, which translates the directed acceleration by the reconnection electric field into thermal particle motion.

Eqn. (2) is analyzed in the same way as (1), by integrating over rectangles of varying sizes. Figure 4 shows that the expectation is correct: the combination of compression, convection, and heat flux serve to reduce the pressure, leading to a negative time derivative. However, much of this reduction is compensated for by the effect of quasi-viscous heating.

Using the symmetry of the pressure tensor, the quasi-viscous contribution in (2) can be written as:

$$H_{qv} = -\frac{2}{3}\sum_{\substack{l,i \\ l \neq i}} P_{li}\frac{\partial}{\partial x_i}v_l = -\frac{2}{3}\left(P_{xz}\frac{\partial}{\partial z}v_x + P_{xy}\frac{\partial}{\partial x}v_y + P_{yz}\frac{\partial}{\partial z}v_y + P_{xz}\frac{\partial}{\partial x}v_z\right) \quad (3)$$

Figure 5 shows that, of these terms, the one proportional to $P_{yze}\frac{\partial}{\partial z}v_{ey}$ dominates by far.

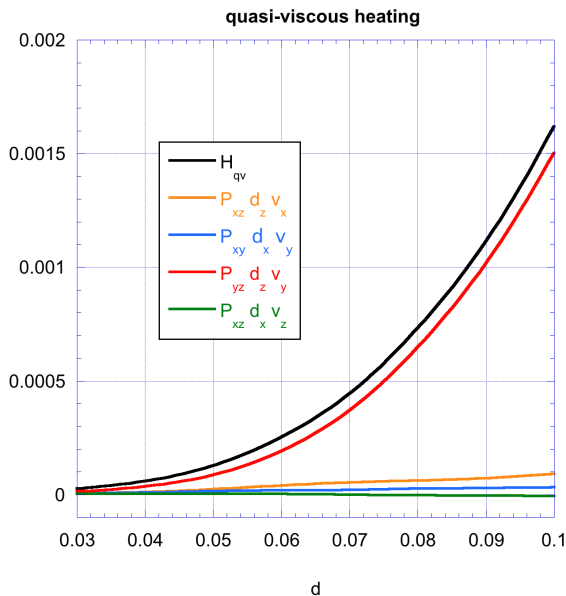

**Figure 5.** Quasi-viscous electron heating. The total of the heating term (3) is shown in black, and the other graphs display the variations of the four terms on the r.h.s. of (3). It is evident that only one term contributes significantly, namely the one proportional to $P_{yze} \frac{\partial}{\partial z} v_{ey}$.

As shown above, the gradient of $P_{yze}$ is also the major contributor to the current reduction (1). Here we see that $P_{yze}$ also provides, combined with a sharp, electron-scale, gradient of $v_{ey}$, the leading order heating term, and it contributes significantly to the overall maintenance of the electron internal energy. This fact supports the intuitive idea that the process, which serves to reduce the current density (and hence the kinetic energy associated with motion of current-carrying electrons), should lead to an increase of the internal energy.

## IV. SUMMARY

The electron internal energy, or pressure, is of substantial importance to the force balance of the diffusion region current layer with the magnetic pressure in the inflow region. Pressure reductions by expansion flows or convection of heated plasma into the outflow region need to be counter-acted so pressure balance can be maintained between the plasma in the current layer and the pressure of the adjacent magnetic field. We see that the diffusion region current layer would likely collapse in the absence of heating effects. This heating is provided by the thermal interaction of current sheet particles with the adjacent magnetic field, leading to quasi-viscous heating effects. While we investigated reconnection in symmetric geometries here, these conclusions apply equally to asymmetric reconnection.

The role of the reconnection electric field is therefore twofold: it is to maintain the current and also to sustain the pressure in the diffusion region. Of those roles current conservation is the most fundamental: in fact, we can now understand that any process reducing the current required by the magnetic field reversal inevitably leads to the establishment of a reconnection electric field. This feature is a surprisingly simple consequence of Ampere's law $\frac{\partial \vec{E}}{\partial t} = c^2 \left( \nabla \times \vec{B} - \mu_0 \vec{j} \right)$. In a Gedankenexperiment, a current density reduction in the electron diffusion region will lead to mismatch between the curl of the magnetic field and the current density. This imbalance immediately generates an electric field, which both reduces the magnetic field gradient requiring the current, and which accelerates charged particles to increase the current density until a match is re-established. This adjustment happens continuously, so that the time derivative of the electric field is very small.

We have thus seen that a reconnection geometry generates a current reduction and heating mechanism by thermal, quasi-viscous, effects. In a reconnecting system, Ampere's law assures that the electric field is continuously regulated to provide the appropriate balance. Therefore, we conclude that the existence of a reconnection electric field can be understood as an electrodynamic consequence of the plasma dynamics inside the inner diffusion region.

The simulation model employed here is translationally invariant, i.e., does not permit fluctuations with k vectors in the out-of-plane direction[20]. The question might therefore arise to which degree such a model represents reality correctly. NASA's

Magnetospheric Multiscale (MMS) mission measures currents, electric fields, and small-scale gradients of plasma quantities in central diffusion zones of near-Earth reconnection[6]. The remarkable match of MMS observations of electron gyroscale features[6,9] – "crescents" with predictions from similar simulations[8,18] suggests that fluctuations are either not strong enough or of frequencies too low to influence electron orbits significantly in the inner diffusion region. A simple reason for this may be that electrons do not remain in this region long enough to interact with the electromagnetic fields of an otherwise unstable mode. The continuous supply of new, uncorrelated particles could suppress instability growth in the inner diffusion region.

Whether the inner diffusion region is always relatively free of fluctuations or not will, without a doubt, be the subject of further investigations, with MMS measurements to provide the ground truth. Here we provide predictions, which can be tested and used as a basis for discrimination of candidate mechanisms. We point out that, in general, any mechanism counteracting the accelerative force of the electric field will also in some form or another have to contribute to the maintenance of the pressure, if only for energy conservation reasons. Finally, our argument regarding the electromagnetic origin of the reconnection electric field will apply irrespective of the underlying current dissipation mechanism.

**ACKNOWLEDGEMENTS**

This work was funded by the NASA MMS project, and in part by grants from NASA-NNX16AG75G, DOE DESC0016278 and NSF AGS1619584, AGS-1202537, AGS-1543598, and AGS-1552142. Work at the University of Bergen was supported by the Research Council of Norway/CoE under contract 223252/F50. The authors recognize the tremendous effort in developing and operating the MMS spacecraft and instruments and sincerely thank all involved.

**APPENDIX: THE ROLE OF THE HEAT FLUX**

It is often assumed that a heat flux vector directed into a certain direction implies that there is motion of particles into that direction, thereby establishing an effective energy transport. However, this assumption can be incorrect. This can be seen rather simply by assuming a one-dimensional system, with a distribution function of the form: $f = n_1 \delta(v - v_1) + n_2 \delta(v - v_2)$. The center of mass velocity is then: $\bar{v} = \dfrac{n_1 v_1 + n_2 v_2}{n_1 + n_2}$.

Evaluating pressure and heat flux (a scalar here) appropriately in the center of mass frame, we obtain: $P = m \dfrac{n_1 n_2}{n_1 + n_2}(v_1 - v_2)^2$ and $Q = m \dfrac{n_1 n_2}{n_1 + n_2}(v_1 - v_2)^3 \left[ n_2^2 - n_1^2 \right]$. It is immediately obvious that choices of densities and velocities can be found, for which $v_1, v_2 > 0$, but $Q < 0$. A somewhat more complex but essentially similar situation applies here.

As an illustration of the underlying distribution function complexity, Figure A1 shows reduced distribution functions $F(v_x, v_z)$ to the left of the diffusion region (x=44, z=0) and to the right (x=47, z=0).

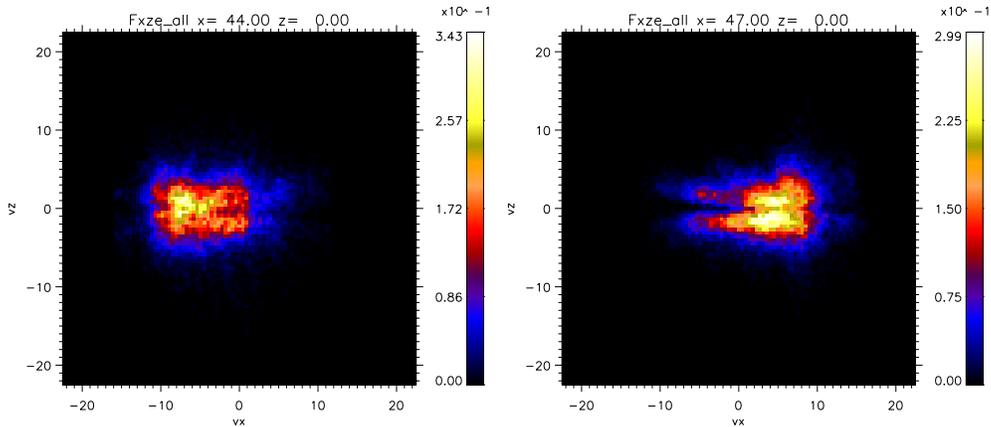

Figure A1: Reduced distributions $F(v_x, v_z)$ in the two outflow regions from the X point. The left panel shows a distribution at (x=44, z=0) and the right panel at (x=47, z=0).

Both distributions demonstrate that the vast majority of particles is moving away from the X point.

In addition to the double-structures associated with populations flowing in from the z direction, it is readily apparent that virtually all particles comprising these distributions are moving away from the reconnection region – yet the heat flux term $q_x = (Q_{xxx} + Q_{xyy} + Q_{xzz})/3.$ contributes positively to the energy balance on the left boundary, and negatively on the right. The total heat flux contribution, as well as the contribution at the individual boundaries of the integration rectangle, are shown below.

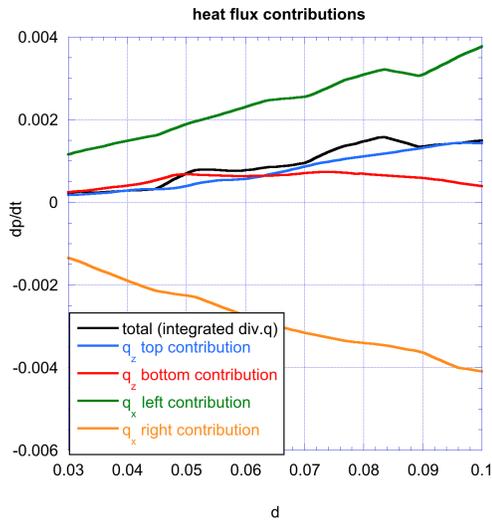

Figure A2: Contributions to the integral of the divergence of the heat flux vector from the different boundaries of the integration box, for varying sizes of the integration box.

On the inflow side, distributions like $F(v_y, v_z)$ above the X point (Figure A3 left, x=45.5, z=0.1) and below (Figure A3 right, x=45.5, z=0.1) show crescent features, major nongyrotropies, and general complexities, which lead to a heat flux $q_z = (Q_{xzz} + Q_{yzz} + Q_{zzz})/3$. directed into the integration volume on both the upper and lower boundaries.

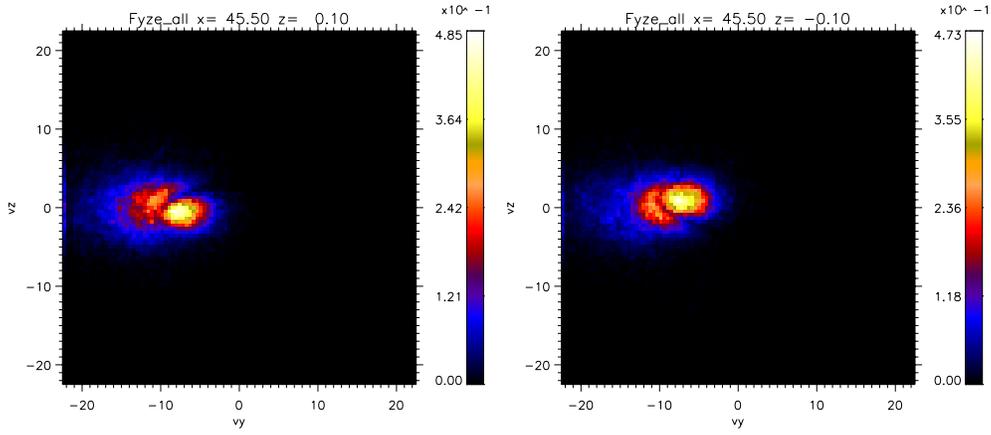

Figure A3: Reduced distributions $F(v_y, v_z)$ in the two inflow regions surrounding the X point. The left panel shows a distribution above the X point (Figure A3 left, x=45.5, z=0.1) and the right panel shows a distribution below (Figure A3 right, x=45.5, z=0.1).

Therefore, we find heat flux is best interpreted as a correction to the transport described by the convection-compression terms in (2) to account for complexity in the particle distributions rather than an energy inflow into the diffusion region. We therefore included it as a correction to convection and compression.